\documentstyle[12pt]{article}

\begin{document}

\title{The Dirac particle on de Sitter background}

\author{Ion I. Cot\u aescu\\ {\it The West University of Timi\c soara,}\\{\it V. 
Parvan Ave. 4, RO-1900 Timi\c soara}}

\maketitle

\begin{abstract}
We show that the Dirac equation on de Sitter background can be analytically 
solved in a special static frame where  the energy eigenspinors can be 
expressed  in terms of  usual angular spinors known from special 
relativity, and a pair of radial wave functions.    

\end{abstract}
\

\section{Introduction}
\

One of the most important but difficult problems of the quantum field theory 
in curved spacetime is the problem of finding analytic solutions of  Dirac 
equation on given backgrounds. Since there are no general methods to obtain 
such solutions, these must 
be derived  in each particular case  separately, starting with a  suitable 
chart and tetrad gauge and by using appropriate calculation procedures. From 
the results reported  up to now \cite{BW,OT,VIL} we understand that the form of 
the solutions obtained  through separation of variables as well as their 
physical significance are  strongly dependent on the choice of all these 
ingredients. In these conditions it is helpful to exploit, in addition, the 
effects of the global symmetries of the background.  

An important case is that of the Dirac equation in spherically symmetric 
(central) static charts which have the global symmetry of the group 
$T(1)\otimes SO(3)$, of time translations and  rotations of the Cartesian space 
coordinates. Recently, we have proposed a Cartesian gauge \cite{C} in which 
the whole theory remains  covariant under this group when  the 
Cartesian holonomic coordinates are used. Then the energy and  angular momentum are 
conserved like in special relativity from which we can take over the method of 
separation of variables in spherical coordinates. In fact, our gauge defines 
Cartesian local (unholonomic) frames which play here the same 
role as the Cartesian natural frame of the Minkowski spacetime, since their 
third axes are just those of projections of the whole angular momentum. This 
allowed us to separate the spherical variables in terms of usual angular spinors 
such that  all the constants involved in  separation of variables get  
physical meaning. Therefore we are sure that the local properties of the Dirac field 
can be correctly interpreted.

On this way we have found a complete formulation of the radial problem of the 
Dirac equation on central static backgrounds, obtaining the radial equations 
and the form of the radial scalar product in the most general case \cite{C}. 
Based on these results, we have solved the Dirac equation on  anti-de Sitter 
static background, giving the formula of the discrete energy levels and the 
form of the energy eigenspinors in the spherical coordinates of a special 
natural frame.  Moreover, we  have observed therein that  the solutions of the 
Dirac equation on de Sitter static backgrounds  could be derived in the same 
manner, following step by step the  technique of the anti-de Sitter case.  This 
is just what we would like to  present here. Our aim is to study the  radial 
problem and its supersymmetry which helps us to  find the radial wave functions 
we need in order to write down the energy eigenspinors in spherical coordinates 
(and natural units, $\hbar=c=1$).

We start in the second section with a short review of our main  results 
concerning the separation of spherical variables in the Dirac equation on 
central static charts. The third section is devoted to the hidden 
supersymmetry of the radial problem  which can be pointed out by using an 
appropriate transformation. This  simplifies the form of the radial 
Hamiltonian  allowing us to completely solve the radial problem, as  shown 
in Sec.4.

\section{Preliminaries}
\

The main point of our approach \cite{C} is the choice of tetrads in Cartesian 
gauge and Cartesian coordinates such that  Dirac equation takes a simple 
form which  transforms manifestly covariant under the rotations of the space 
Cartesian coordinates. Consequently, its theory in the spherical natural 
coordinates, $(t, r, \theta, \phi)$,  can be done  like in special relativity. 
Our tetrad fields  depend  on three  arbitrary functions of $r$, $u$, $v$ and 
$w$, which give the line element 
\begin{equation}\label{(muvw)}
ds^{2}=w^{2}\left[dt^{2}-\frac{dr^2}{u^2}-
\frac{r^2}{v^2}(d\theta^{2}+\sin^{2}\theta d\phi^{2})\right].
\end{equation}
and determine the form of the Dirac equation of the free field, $\psi$, of mass 
$M$. We have shown that this equation has  particular positive frequency 
solutions of energy $E$, 
\begin{eqnarray}
&&\psi_{E,j,\kappa_{j}, m_{j}}(t,r,\theta,\phi)=
u_{E,j,\kappa_{j},m_{j}}(r,\theta,\phi)e^{-iEt}\label{(spin)} \label{(u)}\\
&&=\frac{v}{rw^{3/2}}[f^{(+)}(r)\Phi^{+}_{m_{j},\kappa_{j}}(\theta,\phi)
+f^{(-)}(r)\Phi^{-}_{m_{j},\kappa_{j}}(\theta,\phi)]e^{-iEt}.\nonumber
\end{eqnarray}
Here  $u_{E,j,\kappa_{j},m_{j}}$ are the particle-like energy eigenspinors 
which depend on the radial wave functions $f^{(\pm)}$ and on the four-component 
angular spinors, $\Phi^{\pm}_{m_{j}, \kappa_{j}}$, known from special 
relativity \cite{TH}. These  are orthogonal to each other being completely 
determined by the quantum number $j$, of the  angular momentum, the 
quantum number $m_{j}$, of its projection along the third axis of the local 
Cartesian frame, and the value of $\kappa_{j}=\pm (j+1/2)$. 

The radial wave functions are  solutions of a pair of radial equations 
which can be written in compact form as the eigenvalues problem 
\begin{equation}
H{\cal F}=E{\cal F}
\end{equation}
of the radial Hamiltonian 
\begin{equation}
H=\begin{array}{|cc|}
    Mw& -u\frac{\textstyle d}{\textstyle dr}+\kappa_{j}\frac{\textstyle v}
{\textstyle r}\\
&\\
  u\frac{\textstyle d}{\textstyle dr}+\kappa_{j}\frac{\textstyle v}
{\textstyle r}& -Mw
\end{array}\,,
\end{equation}
in the space of the two-component vectors  ${\cal F}=|f^{(+)}, f^{(-)}|^{T}$ 
where the radial scalar product is 
\begin{equation}\label{(spf)}
({\cal F}_{1},{\cal F}_{2})=\int_{D_{r}}\frac{dr}{u(r)}\, 
{\cal F}_{1}^{+}{\cal F}_{2}\,.
\end{equation}
This selects the "good" radial wave functions (i.e. square integrable functions 
or tempered distributions) which enter in the structure of the particle-like 
energy eigenspinors of (\ref{(spin)}).  Furthermore, the negative frequency 
(antiparticle-like) eigenspinors  can be obtained directly by using the charge 
conjugation \cite{C}.

Here we  use three arbitrary functions ($u$, $v$, and $w$) which define 
the metric and implicitly the form of the radial Hamiltonian. However, it 
is known that, in general, for any central static metric two such functions are 
enough. This means that we have a supplementary degree of freedom allowing us to 
choose suitable radial coordinate.  The most convenient is to 
work in the {\em special} natural frame \cite{C} where, by definition, $u=1$. 
According to (\ref{(muvw)}) its radial coordinate is 
\begin{equation}\label{(spfr)}
r_{s}(r)=\int\frac{dr}{u(r)}+{\rm const}.
\end{equation}
Here the constant assures the condition $r_{s}(0)=0$.   

\section{The supersymmetry of the radial problem}
\

Let us consider  now the problem of the massive Dirac particle on  de Sitter 
background. There exists a static chart with usual spherical coordinates,  
$(t,\hat r, \theta, \phi)$, where  the line element is
\begin{equation}\label{(des)}
ds^{2}=(1-\omega^{2}\hat r^{2})dt^{2}
-\frac{d\hat r^2}{1-\omega^{2}\hat r^{2}}
-\hat r^{2} (d\theta^{2}+\sin^{2}\theta~d\phi^{2}).
\end{equation}
In the corresponding special  frame, denoted fom now by $(t,r,\theta,\phi)$, 
the radial coordinate is    
\begin{equation}
r=\frac{1}{\omega}\,{\rm arctanh}\, \omega \hat r,
\end{equation}
as it is obtained from (\ref{(spfr)}) and (\ref{(des)}). Here,  the radial 
domain is $D_{r}=[0,\infty)$ and  the line element has the form 
\begin{equation}\label{(le)}
ds^{2}=\frac{1}{\cosh^{2}\omega r} \left[dt^{2}-dr^{2}-\frac{1}{\omega^{2}}
\sinh^{2}\omega r~ (d\theta^{2}+\sin^{2}\theta~d\phi^{2})\right]\,,
\end{equation} 
from which  we can identify the functions $v$ and $w$ we need to write the 
radial Hamiltonian 
\begin{equation}
H=\begin{array}{|cc|}
  \frac{\textstyle\omega k}{\textstyle\cosh \omega r}& -\frac{\textstyle d}{\textstyle dr}+
\frac{\textstyle\omega\kappa_{j}}{\textstyle\sinh \omega r}\\
&\\
\frac{\textstyle d}{\textstyle dr}+\frac{\textstyle\omega\kappa_{j}}
{\textstyle\sinh\omega r}
& -\frac{\textstyle\omega k}{\textstyle\cosh \omega r}
\end{array}\,,
\end{equation}
with the notation $k=M/\omega$ (i.e. $Mc^{2}/\hbar\omega$ in usual 
units).

Our radial problem has  hidden supersymmetry like in the anti-de Sitter case 
\cite{C}. This can be easily pointed out with the help of the transformation  
${\cal F}\to \hat{\cal F}=U(r){\cal F}$  where  
\begin{equation}\label{(uder)} 
U(r)=\begin{array}{|cc|}
    \cosh \frac{\textstyle \omega r}{\textstyle 2}&-i\sinh 
\frac{\textstyle \omega r}{\textstyle 2}\\
&\\
    i\sinh \frac{\textstyle \omega r}{\textstyle 2}&\cosh 
\frac{\textstyle \omega r}{\textstyle 2}
\end{array}\,.
\end{equation}
A little calculation shows us that  the transformed  Hamiltonian,   
\begin{equation}\label{(newh)}
\hat H =U(r)HU^{-1}(r)-i\frac{\omega}{2} 1_{2\times 2},
\end{equation}
which gives the new eigenvalue problem
\begin{equation}\label{(trrp)}
\hat H \hat{\cal F}=\left(E-i\frac{\omega}{2}\right)\hat{\cal F},
\end{equation} 
has supersymmetry since it  has  the requested specific form,  
\begin{equation}\label{(ssh)}
\hat H=\begin{array}{|cc|}
    \nu& -\frac{\textstyle d}{\textstyle dr}+W\\
       \frac{\textstyle d}{\textstyle dr}+W& -\nu
\end{array}\,.
\end{equation} 
Here   $\nu=\omega(k-i\kappa_{j})$ is a constant  and 
\begin{equation}\label{(super)}
W(r)=\omega(ik\tanh\omega r + \kappa_{j}\coth \omega r) 
\end{equation}
is the superpotential of the radial problem \cite{C}. The transformed radial 
wave functions  $\hat f^{(\pm)}$ (which are the components of $\hat{\cal F}$) 
satisfy the second order equations resulted from the square of (\ref{(trrp)}). 
These are 
\begin{equation}\label{(2e)}
\left(-\frac{d^2}{dr^2}-\omega^{2}\frac{ik(ik\pm 1)}{\cosh^{2}\omega r}+
\omega^{2}\frac{\kappa_{j}(\kappa_{j}\pm 1)}{\sinh^{2}\omega r}\right)
\hat f^{(\pm)}(r)=
\omega^{2}\epsilon^{2}\hat f^{(\pm)}(r)\label{(od1)},\\
\end{equation}
where we have denoted $\epsilon=E/\omega-i/2$.

\section{Solutions}
\

The solutions of Eqs.(\ref{(2e)})  are well-known to be expressed 
in terms of Gauss  hypergeometric functions \cite{AS},
$F_{\pm}(y)\equiv F(\alpha_{\pm},\beta_{\pm},\gamma_{\pm},y)$,  
depending on the new variable $y=-\sinh^{2}\omega r$, as 
\begin{equation}\label{(gsol)}
\hat f^{(\pm)}(y)=N_{\pm}(1-y)^{p_{\pm}}y^{s_{\pm}}
F_{\pm}(y)
\end{equation}
where 
\begin{equation}
\alpha_{\pm}=s_{\pm}+p_{\pm}+\frac{i\epsilon}{2},\quad
\beta_{\pm}=s_{\pm}+p_{\pm}-\frac{i\epsilon}{2},\quad
\gamma_{\pm}=2s_{\pm}+\frac{1}{2},
\end{equation}
$N_{\pm}$ are normalization factors while the parameters  $p_{\pm}$ and 
$s_{\pm}$ are related with  $k$ and $\kappa_{j}$ through    
\begin{eqnarray}
2s_{\pm}(2s_{\pm}-1)&=&\kappa_{j}(\kappa_{j}\pm 1),\label{(s)}\\ 
2p_{\pm}(2p_{\pm}-1)&=&ik(ik\pm 1).\label{(p)} 
\end{eqnarray}

Furthermore, we have to find the suitable values of these parameters 
such that the functions (\ref{(gsol)}) should be solutions of the transformed 
radial problem (\ref{(trrp)}). If we  replace (\ref{(gsol)}) in (\ref{(trrp)}), after a few manipulation, 
we obtain  
\begin{eqnarray}
&&y(1-y)\frac{dF_{\pm}(y)}{dy}-y\left(p_{\pm}\pm\frac{ik}{2}\right)F_{\pm}(y) 
+(1-y)\left(s_{\pm}\pm\frac{\kappa_{j}}{2}\right)F_{\pm}(y)\nonumber\\
&&=\frac{\eta}{2}\frac{N_{\mp}}{N_{\pm}}\left(\kappa_{j}\pm\frac{1}{2}\pm
i\frac{E\pm M}{\omega}\right)y^{s_{\mp}-s_{\pm}+1/2}(1-y)^{p_{\mp}-p_{\pm}
+1/2}F_{\mp}(y),\label{(idf)} 
\end{eqnarray}
where $\eta=\pm 1$. These equations are nothing else than the usual identities 
of  hypergeometric functions  if the values of $s_{\pm}$, $p_{\pm}$ and 
$N_{+}/N_{-}$ are correctly matched.  First we observe  that the differences 
$s_{+}-s_{-}$ and $p_{+}-p_{-}$ must be  half-integer  since 
we work with analytic functions of $y$. This means that the allowed groups of 
solutions of (\ref{(s)}) are  
\begin{eqnarray}
2s_{+}^{1}=-\kappa_{j}~~~~~ ,&\quad&  2s_{+}^{2}=\kappa_{j}+1,\nonumber\\                            
2s_{-}^{1}=-\kappa_{j}+1 ,&\quad&  2s_{-}^{2}=\kappa_{j},\label{(kka)}
\end{eqnarray}
while  Eq.(\ref{(p)}) gives us
\begin{eqnarray}
2p_{+}^{1}=-ik~~~~~ ,&\quad&  2p_{+}^{2}=ik+1,\nonumber\\                            
2p_{-}^{1}=-ik+1 ,&\quad&  2p_{-}^{2}=ik.\label{(ppa)}
\end{eqnarray}
On the other hand, it is known that the hypergeometric functions of (\ref{(gsol)}) 
are analytical on the domain $D_{y}=(-\infty,0]$, corresponding to $D_{r}$, 
only if  $\Re(\gamma_{\pm})>\Re(\beta_{\pm})>0$ \cite{AS}. Moreover, their 
factors must be  regular on this domain including  $y=0$. Obviously, both these 
conditions are accomplished if we take
\begin{equation}\label{(cond)}
s_{\pm}> 0.  
\end{equation}
We specify that there are no restrictions upon the values of the parameters 
$p_{\pm}$.

We have hence all the possible combinations of parameter values 
giving the solutions of the second order equations (\ref{(2e)}) which satisfy 
the transformed radial problem. These solutions will be denoted  by $(a,b)$, 
$a,b=1,2$, understanding that the corresponding parameters are $s_{\pm}^{a}$,  
$p_{\pm}^{b}$, as given by  (\ref{(kka)}) and (\ref{(ppa)}), and  
\begin{equation}
\alpha_{\pm}^{(a,b)}=s_{\pm}^{a}+p_{\pm}^{b}+\frac{i\epsilon}{2},\quad
\beta_{\pm}^{(a,b)}=s_{\pm}^{a}+p_{\pm}^{b}-\frac{i\epsilon}{2}, \quad
\gamma_{\pm}^{(a)}=2s_{\pm}^{a}+\frac{1}{2}.
\end{equation}
The condition  (\ref{(cond)}) requires to chose $a=1$ when $\kappa_{j}=-j-1/2$, 
and $a=2$  if $\kappa=j+1/2$. Thus, for each given set  $(E,j,\kappa_{j})$ we 
have a pair of different radial solutions, with $b=1,2$. Therefore, it 
is convenient to denote the transformed radial wave functions (\ref{(gsol)}) by 
$\hat f^{(\pm)}_{E,j,a,b}$, bearing in mind that the value of $a$ determines 
that of $\kappa_{j}$. The last step is to calculate the values 
of $N_{+}/N_{-}$. From (\ref{(idf)}) it results 
\begin{eqnarray}
\kappa_{j}=-j-\frac{1}{2}:   &&\eta\frac{N_{-}^{(1,1)}}{N_{+}^{(1,1)}}=
-\frac{\alpha_{+}^{(1,1)}}{\gamma_{+}^{(1)}}, \qquad 
~~~\eta\frac{N_{-}^{(1,2)}}{N_{+}^{(1,2)}}=
\frac{\beta_{+}^{(1,2)}}{\gamma_{+}^{(1)}}-1, \\ 
\kappa_{j}=j+\frac{1}{2}:   &&\eta\frac{N_{+}^{(2,1)}}{N_{-}^{(2,1)}}=
1-\frac{\beta_{-}^{(2,1)}}{\gamma_{-}^{(2)}}, \qquad 
\eta\frac{N_{+}^{(2,2)}}{N_{-}^{(2,2)}}=
\frac{\alpha_{-}^{(2,2)}}{\gamma_{-}^{(2)}}.  
\end{eqnarray}
Notice that here  $\gamma_{+}^{(1)}= \gamma_{-}^{(2)}=j+1$.  

Now we can restore the form of the original radial wave functions of 
(\ref{(u)}) by using the inverse of (\ref{(uder)}). In our new notation these 
wave functions are 
\begin{equation}\label{(ff)}
f^{(\pm)}_{E,j,a,b}(r)=\cosh\frac{\omega r}{2}\hat f^{(\pm)}_{E,j,a,b}(r)
\pm i\sinh\frac{\omega r}{2}\hat f^{(\mp)}_{E,j,a,b}(r).
\end{equation}
For very large $r$ (when $y\to -\infty$) the hypergeometric functions 
behave as $F_{\pm}(y)\sim (-y)^{-\alpha_{\pm}}$ \cite{AS}. Thereby we  
deduce that $\hat f^{(\pm)}_{E,j,a,b}\sim \exp(-i\epsilon \omega r)$ and   
\begin{equation}\label{(ffa)}
f^{(\pm)}_{E,j,a,b}\sim e^{-iEr}.
\end{equation}  
This means that the functions (\ref{(ff)}) represent tempered distributions 
corresponding to a continuous energy spectrum. On the other hand, (\ref{(ffa)}) 
indicates that this  energy spectrum covers  the whole real axis, as it seems 
to be natural since the  metric is not asymptotically flat. However, in our 
opinion, it is premature to draw definitive conclusions before to carefully 
study the properties of the radial state space.  Anyway, it is clear that the 
energy spectrum is continuous, without discrete part, while the energy levels 
are infinitely degenerated since there are no restrictions upon the values of 
$j$, which can be any positive half-integer. 

\section{Comments}
\

Here we have derived the  solutions of the Dirac equation on  de Sitter  
background. This was possible grace of our general method based on the 
Cartesian tetrad gauge  which preserves the maximal global symmetry 
of the central static charts. We must say that these solutions are 
different from the other ones obtained by using also Cartesian gauges but in 
the spherical coordinates of  moving charts \cite{VIL}. The argument is that  
our solutions rewritten in the comoving frame have no more separated variables.

>From the technical point of view, we have used here the same procedure as in 
Ref.\cite{C}. Thus it is clear that the Dirac equation on  de-Sitter or
anti-de Sitter backgrounds can be solved in the same manner. However, despite 
this fact, the results are different since the corresponding energy spectra 
are of different kinds. We believe that these are good examples of free Dirac 
fields with continuous and respectively discrete energy spectra which could 
help us to understand  some sensitive aspects of the quantum theory in curved 
spacetime. We refer especially to the structure of the operator algebra and its 
dependence on the choice of the tetrad gauge in a given chart. A guide in this 
direction could be our recent study of the algebras of the one-dimensional 
relativistic oscillators \cite{CO} which have similar superpotentials  
as those of the transformed  Hamiltonians of the de Sitter or anti-de Sitter 
radial problems.


\begin{thebibliography}{20}

\bibitem{BW} 
D. R. Brill and J. A. Wheeler, {\it Rev. Mod. Phys.} {\bf 29}, 465 (1957);  

\bibitem{OT}
V. S. Otchik, {\it Class. Quantum Grav.} {\bf 2}, 539 (1985)

\bibitem{VIL}
G. V. Shishkin and V. M. Villalba, {\it J. Math. Phys.} {\bf 30}, 2132 (1989);
{\it J. Math. Phys.} {\bf 33}, 2093 (1992);
V. M. Villalba and U. Percoco, {\it J. Math. Phys.} {\bf 31}, 715 (1990);
G. V. Shishkin, {\it Class. Quantum Grav.} {\bf 8}, 175 (1991);


\bibitem{C}
I. I. Cot\u aescu, preprint gr-qc/9803042 (to appear in 
{\it Int. J. Mod. Phys. A})

\bibitem{TH} 
B. Thaller,  {\it The Dirac Equation}, Springer Verlag, Berlin 
Heidelberg, 1992


\bibitem{AS}
M. Abramowitz and I. A. Stegun, {\it Handbook of Mathematical Functions}
(Dover, 1964)



\bibitem{CO}
I. I. Cot\u aescu and G. Drag\u anescu, {\it J. Math. Phys.} {\bf 38}, 5505 (1997);
I. I. Cot\u aescu, {\it J. Math. Phys.} {\bf 39}, 3043 (1998); {\it Theoret. 
Math. Comp. Phys.\footnote{New series of {\it Annals of the West University of 
Timis\c oara} (ISSN: 1453-9225)}} {\bf 1}, 15 (1998) 

 
\end{thebibliography}
\end{document}